# Dirac Strings and Magnetic Monopoles in Spin Ice $Dy_2Ti_2O_7$

D.J.P. Morris[1], D.A. Tennant[1,2], S.A. Grigera[3,4], B. Klemke[1,2], C. Castelnovo[5], R. Moessner[6], C. Czternasty[1], M. Meissner[1], K.C. Rule[1], J.-U. Hoffmann[1], K. Kiefer[1], S. Gerischer[1], D. Slobinsky[3], and R.S. Perry[7]

[1]*Helmholtz-Zentrum Berlin für Materialien und Energie, Glienicker Str. 100, Berlin D-14109, Germany*

[2]*Institut für Festkörperphysik, Technische Universität Berlin, Hardenbergstr. 36, Berlin D-10623, Germany*

[3]*School of Physics and Astronomy, North Haugh, St Andrews, Fife KY15 9SS, United Kingdom*

[4]*Instituto de Física de Líquidos y Sistemas Biológicos, CONICET, UNLP, La Plata, Argentina*

[5]*Rudolf Peierls Centre for Theoretical Physics, 1 Keble Road, Oxford OX1 3NP, United Kingdom*

[6]*Max Planck Institut für Physik Komplexer Systeme, Nöthnitzer Str. 38, D-01187 Dresden, Germany*

[7]*School of Physics, University of Edinburgh, Mayfield Road, Edinburgh EH9 3JZ, United Kingdom*

*To whom correspondence should be addressed. Email: D.J.P.M (jonathan.morris@helmholtz-berlin.de), D.A.T (tennant@helmholtz-berlin.de) or S.A.G (sag2@st-and.ac.uk).*

**While sources of magnetic fields — magnetic monopoles — have so far proven elusive as elementary particles, several scenarios have been proposed recently in condensed matter physics of emergent quasiparticles resembling monopoles. A particularly simple proposition pertains to spin ice on the highly frustrated pyrochlore lattice. The spin ice state is argued to be well-described by networks of aligned dipoles resembling solenoidal tubes – classical, and observable,**

**versions of a Dirac string. Where these tubes end, the resulting defect looks like a magnetic monopole. We demonstrate, by diffuse neutron scattering, the presence of such strings in the spin-ice $Dy_2Ti_2O_7$. This is achieved by applying a symmetry-breaking magnetic field with which we can manipulate density and orientation of the strings. In turn, heat capacity is described by a gas of magnetic monopoles interacting via a magnetic Coulomb interaction.**

Despite searching within the cosmic radiation, particle colliders and lunar dust, free magnetic monopoles have not been observed (*1,2*). This is particularly disappointing as unification theories have predicted their existence. Dirac's original vision for monopoles involves a string singularity carrying magnetic flux the ends of which act as north and south monopoles. We report the observation of analogous strings and magnetic monopoles in spin ice, magnetic compound $Dy_2Ti_2O_7$ with a pyrochlore lattice structure. This is a realisation of magnetic fractionalisation in 3D, a separation of north and south monopoles.

Dysprosium Titanate contains magnetic $^{162}Dy$ ions in the highly frustrated pyrochlore lattice which have ferromagnetic exchange and dipolar interactions between the spins. The pyrochlore lattice is a three dimensional structure built from corner sharing tetrahedra (Fig. 1A). Spin-ice is realised on this lattice when spins placed on the vertices are constrained to point radially into or out of the tetrahedra, and are coupled ferromagnetically, or, as in the case of $Dy_2Ti_2O_7$, through dipolar coupling (*3*). This leads to the lowest energy spin configurations obeying the "ice rules" of two spins pointing into, and two out of, each tetrahedron. This is equivalent to the physics of the proton arrangement in ice, where two protons sit close to each oxygen and two far away – and indeed spin ice exhibits the Pauling ice entropy $S \approx \frac{R}{2}\ln\frac{3}{2}$ per spin (*4,5*), reflecting a huge low-energy density of states in zero magnetic field.

Each spin can be thought of as a small dipole or solenoid channelling magnetic flux into and out of a tetrahedron. The ice-rules are too weak to impose magnetic long-range order, but they do induce dipolar power-law correlations resulting in characteristic pinch-point features in neutron scattering (*6-10*).

A spin flip violates the ice rule in two tetrahedra, at a cost of about 2 K per tetrahedron in $Dy_2Ti_2O_7$. It was proposed that to a good approximation this can be viewed as the formation of a pair of monopoles of opposite sign in adjacent tetrahedra (*11*). Crucially, these monopoles are deconfined (Fig. 1A), they can separate and move essentially independently. Thus, the equilibrium defect density is determined not by the cost of a spin flip but by the properties of the gas of interacting monopoles. In Fig. 1B, we compare the measured heat capacity to Debye-Huckel theory (*12*), which describes a gas of monopoles with Coulomb interactions. This theory is appropriate to low temperatures, where the monopoles are sparse, and it captures the heat capacity quantitatively. At higher temperatures, spin ice turns into a more conventional paramagnet and the monopole description breaks down (see supporting online material). Together with a recent analysis of dynamic susceptibility (*13*), this lends strong support to the monopole picture of the low-temperature phase of spin ice.

Monopole deconfinement is reflected in the spin configurations: as the two monopoles of opposite sign separate, they leave a tensionless string of reversed spins connecting them. These strings of reversed flux between the monopoles can be viewed as a classical analogue of a Dirac string: in the theory of Dirac (*14*), these are infinitely narrow, unobservable solenoidal tubes carrying magnetic flux density (**B**-field) emanating from the monopoles. Here, the strings are real and observable thanks to the preformed dipoles of the spins; strings can change length and shape, at no cost in energy other than the magnetic Coulomb interaction between their endpoints.

As the strings consist of magnetic dipoles, the method of choice for imaging them is magnetic neutron scattering. As a first step, using diffuse neutron scattering techniques we have measured three-dimensional correlation functions in $Dy_2Ti_2O_7$. Fig. 2A shows the results with no applied magnetic field. One important property of the correlation functions in zero-field is the existence of pinch points, a signature of the spin ice state, which can be seen in the experiment as 3D singularities (Fig. S1). In order to compare with the neutron scattering measurements, we performed a large-N (self-consistent mean-field) calculation (*6*,*15*), supplemented with the relevant geometric factors for neutron scattering experiments and the magnetic form factor of Dy (Fig. 2B). Good agreement between theory and experiment (see also Ref. *16*) demonstrates that the correlations do indeed follow the predicted dipolar form, in spite of the fact that direct observation of the pinch-points is not possible as they are covered by Bragg peaks.

In the Dirac string picture, the spin-ice ground state satisfying the ice rules can be considered as a dense network of inter-woven strings that are either closed (i.e. loops) or terminate at the surface of the sample. If the applied field is zero, the strings have no privileged orientation and they describe isotropic, intertwined 3D random walks of arbitrary length. Excitations correspond to monopoles at the end of Dirac strings (broken loops) in the bulk. The cost of lengthening such Dirac strings is solely against the weak attractive force between the monopoles at their ends (*11*). Indeed, as the strings fluctuate between different configurations consistent with a given distribution of monopoles, there is not even a unique way of tracing their paths.

However, there exists an elegant remedy: the application of a large magnetic field along one of the principal axes (here we choose the [001] direction) orients all spins. The resulting ground state is unique and free of monopoles – the ice rules are observed everywhere, and each tetrahedron is magnetised in the [001] direction. Upon lowering the field, sparse strings of flipped spins appear against the background of this fully magnetised ground state. In the absence of monopoles, such strings must span the length of the sample and terminate on the surface; otherwise, they can terminate on magnetic monopoles in the bulk (as explained above).

The presence or absence of the strings at a given temperature and field is determined by a balance between the energy cost of producing them and the gain in entropy due to their presence. As pointed out in Refs. *17* and *18* each link in the string will involve a spin being reversed against the field (note that this still maintains the two-in-two-out ice rules along the string). Each spin flip costs a Zeeman energy of $\frac{2}{\sqrt{3}}h$, with $h = gmB$ the strength of the field applied along [001], $g$ is the Landé g-factor, $m = 10\mu_B$ being the magnetic moment per Dy ion, and $B$ is the applied field. As there are two possibilities to choose for the continuation of the string, there is an associated entropy per link of $s = k_B \ln 2$. The free energy per link, as the string becomes large, is $f = u - Ts = \frac{2}{\sqrt{3}}h - k_B T \ln 2$. For fields above $h_K = \frac{k_B T \ln 2}{2/\sqrt{3}}$, the number of strings goes to zero as the free energy of formation is macroscopic and positive. However at the field $h_K$ a transition occurs where strings spontaneously form as the free energy becomes favourable and the entropy of string formation wins. This transition is a three-dimensional example of a Kasteleyn transition (*17*), a highly unusual topological phase transition. Thus, by measuring close to this transition, we can dial up a regime where the strings are sparse and oriented against the field direction. This allows us to check, qualitatively and quantitatively, the properties of these strings: in particular, we will find that they lead to a qualitative signature in the neutron scattering, which can be manipulated by tilting the field. In the following, we first locate the Kasteleyn transition and hence $h_K$, then we discuss the neutron scattering data in detail by comparing it to a theoretical model. Evidence of this transition can be seen in the magnetisation as a function of temperature and field along [001] (Fig. 3A). Whilst the magnetisation in response to field changes at constant temperature only equilibrates well above 0.6K, spin ice can remain closer to its thermodynamic equilibrium behaviour when cooled in zero magnetic field. Above this temperatures the system shows a transition to saturation at a field $h_S(T)$ consistent with that expected for the three dimensional Kasteleyn transition. Indeed, in the ergodic region of the phase diagram the saturation field $h_S$ coincides with the Kasteleyn field $h_K$, where a kink in the magnetisation appears as it reaches its saturation value (*17*). However, below around 0.6 K, equilibration times become so long that the system starts freezing (*17*), and the measured magnetisation is no longer an equilibrium property. Another signature of this freezing, as shown in the right hand side of Fig. 3A, is that the saturation field $h_S$ (dotted white line) becomes temperature independent. Our observations on freezing in thermodynamic and transport properties will be published elsewhere.

Here we restrict ourselves to equilibrium phenomena: all the neutron measurements in this study were undertaken above 0.6 K. These experimental temperatures are high enough compared to the creation energy of monopoles that the assumption of perfect compliance to the ice-rules is no longer valid and the transition is rounded: the (small) thermal density of monopoles leads to strings of finite length.

The panels in Fig. 3B show reciprocal space slices at a field near saturation of $h = 5/7\ h_S$. Instead of the two lobes coming down to a pinch point in zero-field, cone-like scattering emanates from what was the position of the pinch point. As the field is decreased the diffuse scattering smoothly deforms back to the zero field form.

As described above, the strings execute a random walk; when their density is small, interactions between them can be neglected to a first approximation, so that the spin correlations are those of a diffusion process with the z coordinate assuming the role usually played by time:

$$C(x, y, z) \approx \frac{1}{z} \exp\left( \gamma \frac{x^2 + y^2}{z} \right)$$ (see supporting online material).

In order to capture lattice effects (and to fix the constant $\gamma$), we simulate random walks on the pyrochlore lattice. The correlations we find are essentially unchanged if we include interactions in the form of hardcore exclusion (see supporting online material). Because there is a finite thermal population of monopoles and defects in the material, we expect the strings to be finite in length. For the field of $h = 5/7\, h_S$ and 0.7 K a string length of the order 50 sites is required for agreement with the data. Indeed, for this temperature in zero field, the density of monopoles in numerical simulations is found to be very low – well below 1 % of all tetrahedral – in keeping with a large string length.

The scattering from a large ensemble of such hard-core walks has been calculated including all the geometrical factors for the neutron scattering cross section. As can be seen from the side-by-side comparison of the data and modelling, the string configurations account very well for the data and reproduce the cone of scattering observed.

We repeated the experiment with an effective field tilted $\approx 10°$ towards the [011] direction, in order to induce a net tilt in the meandering of the strings. The cone of diffuse scattering collapses into sheets of scattering at an angle of $45°$, matching the opening angle of the original cone. This sharp sheet in reciprocal space widens with decreasing field (Fig. 4C). Within the random walk model, tilting the applied magnetic field changes the relative probabilities of each step (thus generating a biased random walk) as inequivalent spin flips incur different energy costs. The field and temperature dependent Boltzmann factor for the ratio of probabilities of stepping from tetrahedral site 1 to site 3, or site 1 to site 4 is

$$\frac{p(1 \to 3)}{p(1 \to 4)} = \exp\left(\frac{2\mu|h|(\cos\theta_1 - \cos\theta_2)}{k_B T}\right) \qquad (1)$$

Here $\theta_1$ and $\theta_2$ are the angles between the magnetic moments on the two final sites and the magnetic field $h$. Because of the ferromagnetism induced in spin ice, demagnetisation effects must be carefully accounted for in the modelling. A bias of 0.8:0.2 at $^4/_7 h_S$, and 0.64:0.36 at $^2/_7 h_S$ is anticipated from Eqn. 1. Using these weighting factors and modelling the new ensemble of Dirac strings, the tilts and widths of the scattering are well reproduced.

There still remains the issue of inter-string correlations. Comparison with the hard-core string model (Figs. 4C and 4D), shows that the simple random walk approach does not capture the intensity distribution so well. Figure 4D shows a cut through the walls of scattering in the ($h,k,2\eta+k$) plane where $\eta$ is an integer. The intensity within the sheets is indicative of correlations between strings. This may indicate short range ordering of the strings, and the increased intensities around (0, $^2/_3$, 2 $^2/_3$) would suggest a local hexagonal patterning. Further calculations are needed including interactions between strings to clarify this in detail.

Our study of the spin-ice state in zero field and under an applied magnetic field along [001] lend support to the strongly correlated and degenerate, nature of the ground state, and the resulting long range dipolar correlations. The low energy excitations of such a complex ground state are remarkable in their simplicity, and can be accounted for, to a very good extent, by weakly interacting pointlike quasiparticles, the magnetic monopoles, connected by extended objects, the Dirac strings of reversed spins. In zero-field, a description based on a gas of monopoles accounts for the measured low temperature specific heat. Under fields applied along [001], the picture is that of Dirac strings of reversed spins meandering along the direction of the applied field and terminating on monopoles. This picture accounts very well for the spin correlations observed through neutron scattering. The behaviour of the magnetisation as a function of temperature and field near saturation, where all the strings are expelled, is an example of a (thermally rounded) three-dimensional Kasteleyn transition. This description is rather robust, and gives a simple picture of the spin configuration under tilted fields in terms of biases in the string direction. Our main result consists in the experimental identification of these string-like spin excitations in a gas of magnetic monopoles. These constitute hardy and practical building blocks for the understanding of the low energy behaviour of spin ice. Perhaps the most intriguing open issue is the precise connection between these building blocks and the low-temperature freezing observed in the spin ice compounds (*13*,*19*).

Our work presents to our knowledge the first direct evidence of Dirac strings. It provides compelling evidence for dissociation of north and south poles – the splitting of the dipole - and the identification of spin ice as the first fractionalised magnet in three dimensions. The emergence of such striking states is profoundly important in physics both as a manifestation of new and singular properties of matter and routes to potential technologies. Examples of fractionalisation are extremely rare and nearly all confined to one and two dimensions, and so the three dimensional pyrochlore lattice provides us with a promising new direction for future exploration both in magnets and exotic metals.

Our findings are of relevance not only from a fundamental physics aspect – we have evidenced a new set of quasiparticles which have no elementary cousins – but also because it initiates the study of a new type of degree of freedom in magnetism, namely an object with both local (pointlike monopole) and extended (tensionless Dirac string) properties. $Dy_2Ti_2O_7$ is an exceptionally clean material, and with the full array of powerful experimental techniques and pulsed fields, equilibrium and non-equilibrium properties can be comprehensively addressed which will present a substantial statistical physics and dynamical systems challenge. This may throw light on other systems where string like objects can appear – for instance in the study of polymers or nanoclusters – but where freezing of solvents and inhomogeneities can restrict access to all the physics. Spin ice then is a remarkable material which promises to open up new and complementary insights on both emergence of fractionalised states and the physics of ensembles of strings in and out of equilibrium.

20. We wish to thank Shivaji Sondhi for help and encouragement, John Chalker for insights into the expected behaviour in field of the correlation functions, and Konrad Siemensmeyer with his help with sample cutting and preparation. Jochen Heinrich helped with COMSOL Multiphysics used for the demagnetisation calculations. S.A.G. would like to acknowledge financial support from the Royal Society (UK); C.Ca. was supported by EPSRC under Grant No. GR/R83712/01.

Fig. 1

A)

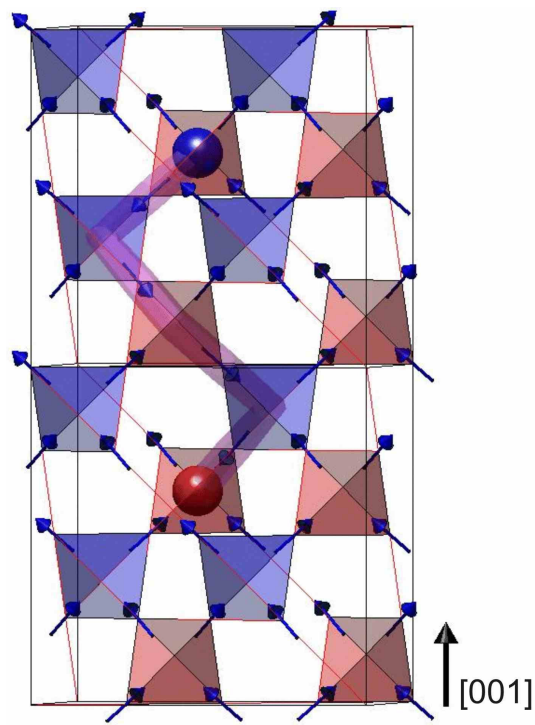

B)

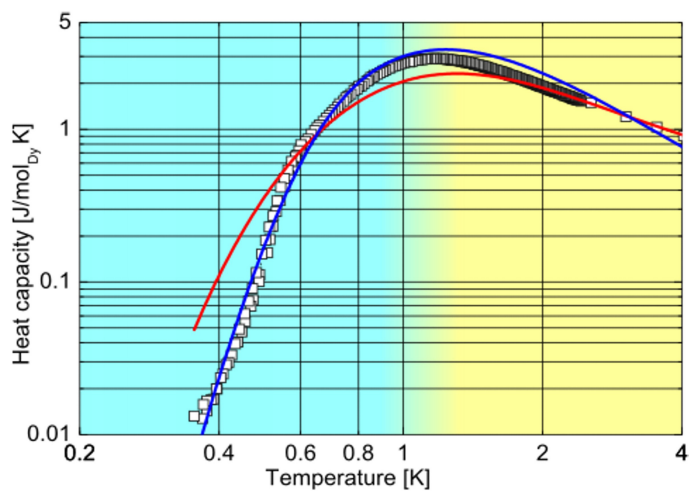

Figure 1. Gas of deconfined magnetic monopoles. A) The Ising spins are constrained to point along the direction connecting the centres of the tetrahedra they belong to. The lowest energy for a tetrahedron is obtained for a two-in-two-out configuration, as illustrated. There are six such configurations with net ferromagnetic moments along one of the six equivalent <100> directions. The non-collinearity of the Ising axes is the source of the frustration in spin ice. In $Dy_2Ti_2O_7$ the 'Ising' crystal field doublet is separated from other levels by more than 100K. Applying a field, $\boldsymbol{B} \parallel [001]$, results in a preference for aligning the tetrahedral magnetisation with the applied field direction. In the three dimensional pyrochlore, Dirac strings of flipped spins terminate on tetrahedra where magnetic monopoles reside. B) The measured heat capacity per mole of $Dy_2Ti_2O_7$ at zero field (open squares) is compared with a Debye-Huckel theory for the monopoles (blue line) and the best fit to a single-tetrahedron (Bethe lattice) approximation (red line). The ice blue (yellow) backgrounds indicate the spin ice (paramagnetic) regimes, respectively.

Fig. 2

A) 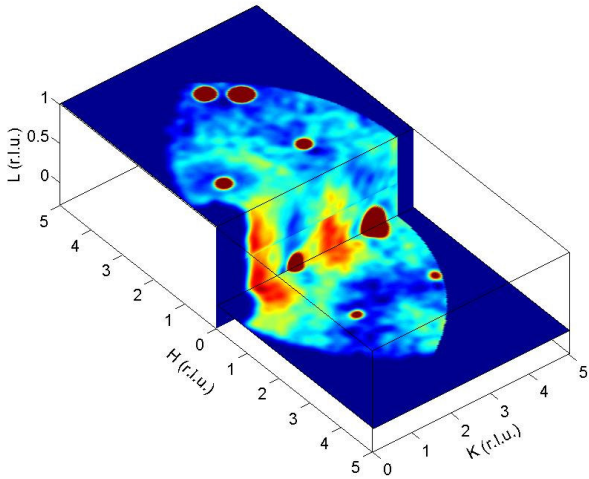

B) 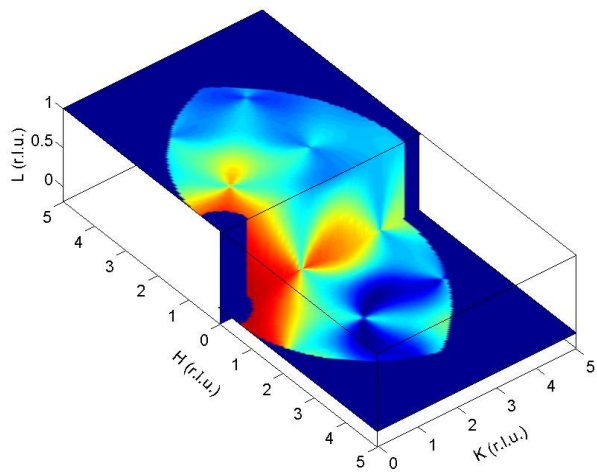

Figure 2. Three dimensional rendering of the dipolar correlations in reciprocal space (*hkl*) of spin ice at 0.7 K. A) Neutron diffraction data taken at 0.7 K and 0.0 T on E2, HZB, with the diffuse peaks at (030) and minima in ($^3/_2$ $^5/_2$ 0) positions. Bragg peaks (red spots) lie on top of the pinch points. A secondary Bragg peak is from a smaller crystallite. B) Pinch points are found in the correlation functions and these are three dimensional in nature, with the diffuse scattering constricting at the reciprocal lattice point (020) and its equivalents. Diffuse peak and scattering minima positions agree with the data.

Fig. 3

A)

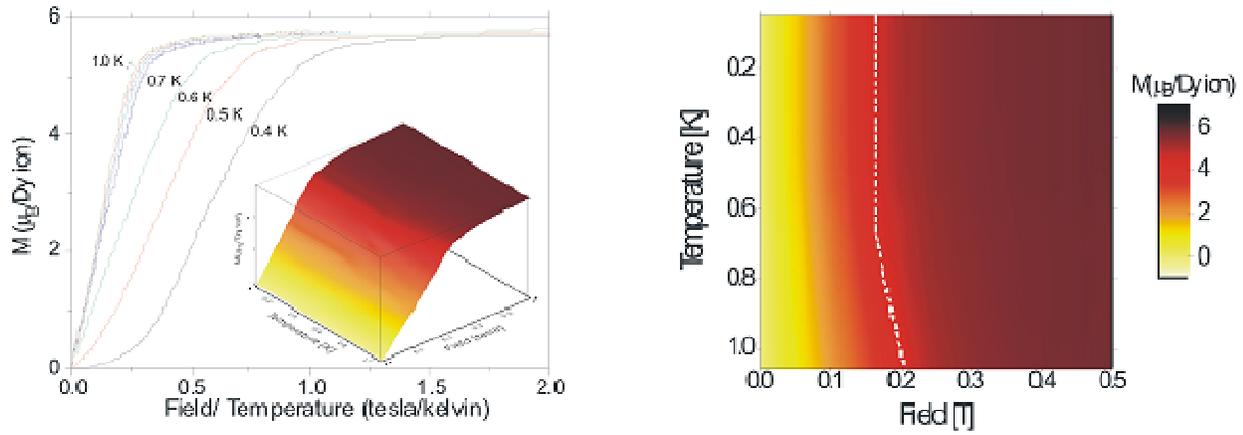

B)                                    C)

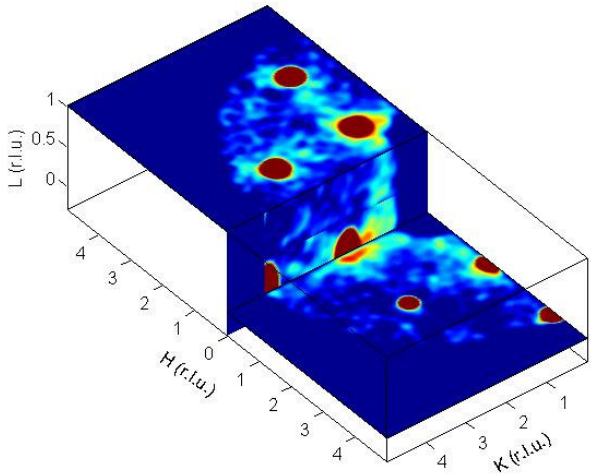 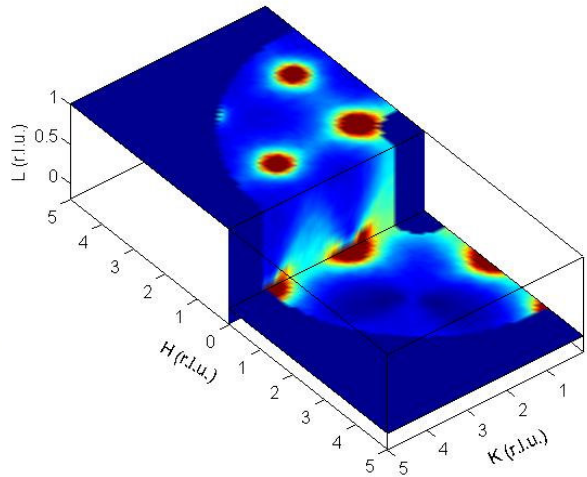

Figure 3. Magnetisation and diffuse neutron scattering with field applied along the [001]. A) Left panel, magnetisation plotted versus [001] field over temperature for fixed temperatures of 0.4, 0.5, 0.6, 0.7, 0.85, 0.9 and 1.0 K, showing a clear departure from $h/T$ scaling at 0.6 K and below. Inset: Surface of magnetisation as a function of temperature and field as constructed from over 70 field and temperature sweeps. Right panel, contour plot of the magnetisation as a function of field and temperature. The dotted white line shows $h_S$ which is seen to freeze below 0.6 K. B) 3D representation of the single-

crystal neutron diffraction data from E2, HZB, at $^5/_7 h_S$ and 0.7 K showing a cone of scattering coming from (020) Bragg peak. C) Calculation of diffuse scattering characteristic of the weakly biased random walk correlations with bias of 0.53:0.47 and $\boldsymbol{B}_{int} \parallel [001]$.

Fig. 4

A) 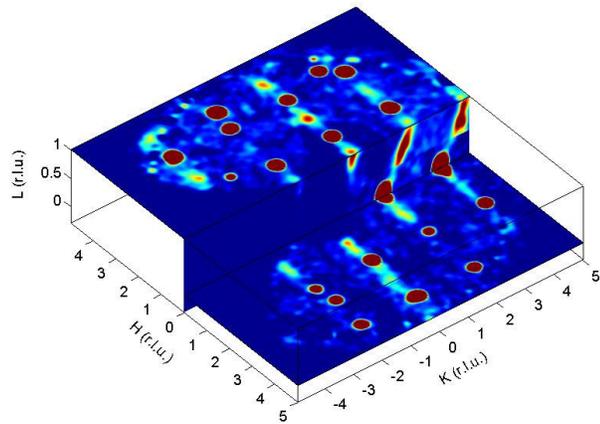
B) 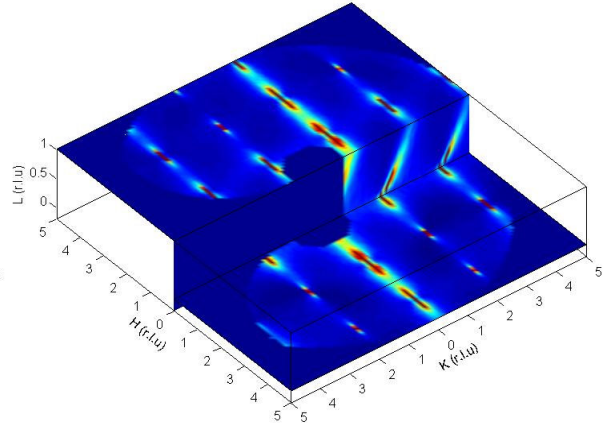

C) 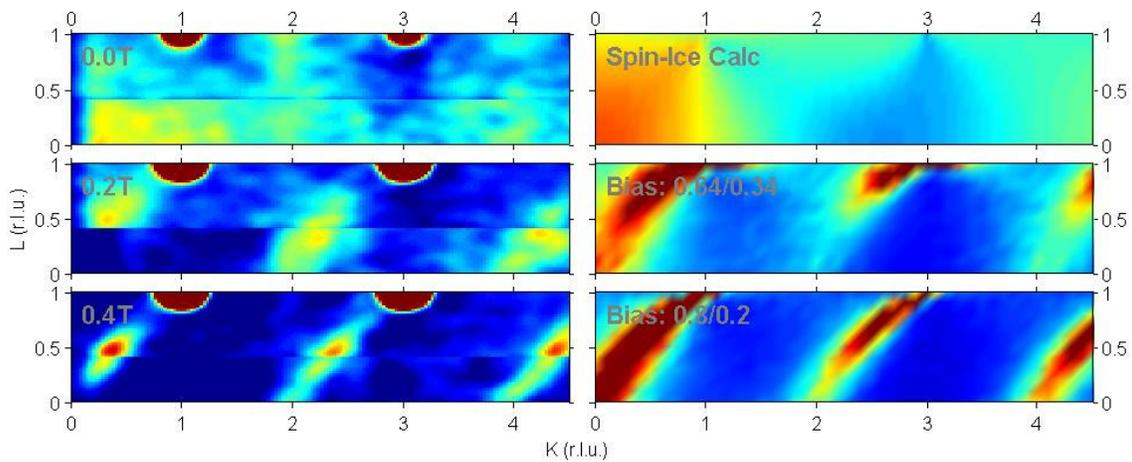

D) 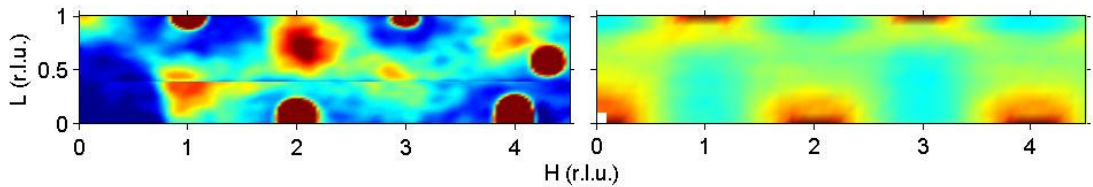

Figure 4. Biased random walks in tilted field. A) Neutron diffraction data from E2, HZB showing the (*hk*0), (*hk*1) and (0*kl*) planes taken at 0.7 K and a field of $^4/_7 h_S$. The red spots are Bragg peaks, with the peaks at (2.7,-1.8,0), (1.3,-2.3,1), (3.8,-0.9,1) and (3.5,2.5,1) being from a smaller second crystallite. B) Random-walk string model with biasing 0.8:0.2. C) Field dependence of the diffuse scattering and calculations in the (1,*k*,*l*) plane. Spin-ice scattering collapses into walls of scattering at higher fields. Further understanding of the Dirac string and monopole physics may improve our modelling of the data. D) Data versus calculation for the (*h*,*k*,2η+*k*) diffuse wall of scattering, where η is an integer (here η=0).

# Supporting online material

**Tensionless Dirac Strings and Magnetic Monopoles in Spin Ice $Dy_2Ti_2O_7$**

D.J.P. Morris, D.A. Tennant, S.A. Grigera, B. Klemke, C. Castelnovo, R. Moessner, C. Czternasty, M. Meissner, K.C. Rule, J.-U. Hoffmann, K. Kiefer, S. Gerischer, D. Slobinsky, and R.S. Perry

**Methods**

Neutron diffraction measurements have been taken on the flat cone diffractometer E2, at Helmholtz-Zentrum Berlin (HZB) in Germany, down to dilution temperatures and with applied magnetic field along the [001] crystallographic direction. Complementary heat capacity (HZB) and magnetisation data (St. Andrews University) confirm the dipolar nature of the low-temperature correlations and support the picture of monopole excitations and Dirac strings (Oxford, MPI Dresden).

Isotopically enriched Dy-162 is used for all the studies reported here as it has no nuclear moment, and reduced neutron absorption cross section. The absence of coupling to the nuclear moment is important in providing a pristine example of a true spin ice system where non-ergodic effects turn out to be of considerable importance

**Neutron measurements**

The neutron scattering measurements were undertaken using the E2 flat cone diffractometer at the BER II research reactor, Helmholtz-Zentrum Berlin. A PG monochromator provided a wavelength $\lambda$=2.39 Å, and the diffuse scattering was measured with four 30x30cm$^2$ position sensitive detectors. A single crystal sample of $Dy_2Ti_2O_7$ with 95-98% Dy-162 was aligned with the [*hkl*]=[100] and [010] directions in the horizontal scattering plane and a cryomagnet with a field along the [001] *c*-direction provided magnetic fields of up to 4 Tesla. A dilution insert was used to reach temperatures down to 0.04 K although due to spins freezing all measurements were at 0.7 K. The freezing explains the discrepancy between our neutron results and those in previous publications (*S1*,*S2*). The data shown in each E2 panel includes two datasets covering below *l*=0 r.l.u. and above *l*=1 r.l.u. measured with flat-cone angle of 0° and 8° respectively. Data were collected then converted into 3D reciprocal space using the instrument's TVnexus software. A measurement at saturated field was subtracted (with Bragg peaks in the background being replaced by a suitable region of the scattering in ***Q***-space) to reduce background contamination of the diffuse scattering.

Pinch points have been observed with applied field along the [111] direction as previously reported by other groups (*S3*). Below is an example of such a point.

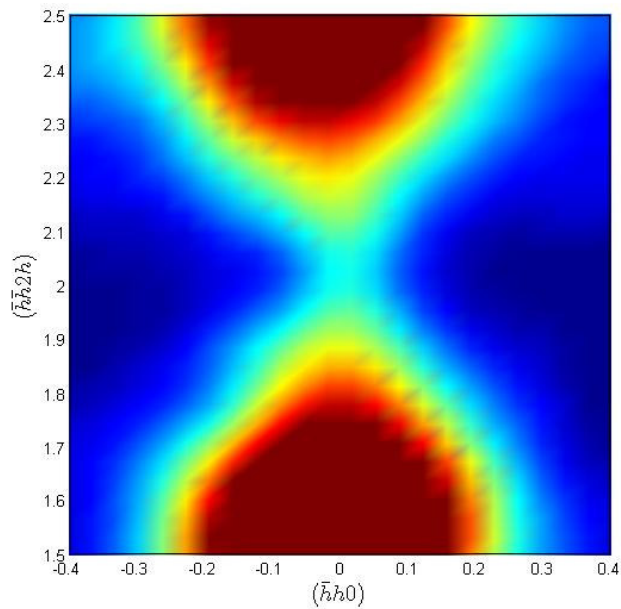

Fig. S1. Pinch point in E2 neutron diffraction data with the magnetic field parallel to the [111] direction.

The cone of scattering for the field parallel to [001] is shown more clearly below. At L=0.2 the circle made by bisecting the cone can be seen at (20L) in Fig. S4. The Bragg peak is missing here since the sample was slightly misaligned, to within a degree or two, as the plot shows by the mismatch in the Bragg peaks along the axes. At L=0.3 the circle is obvious around (20L) and (02L).

| E2 Data | Random walk model |
|---|---|

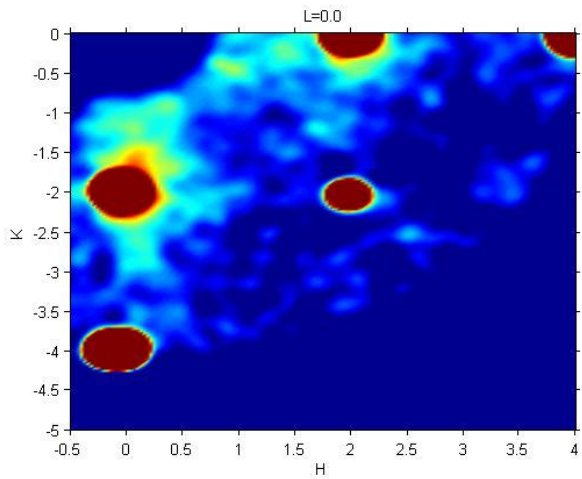

S2.

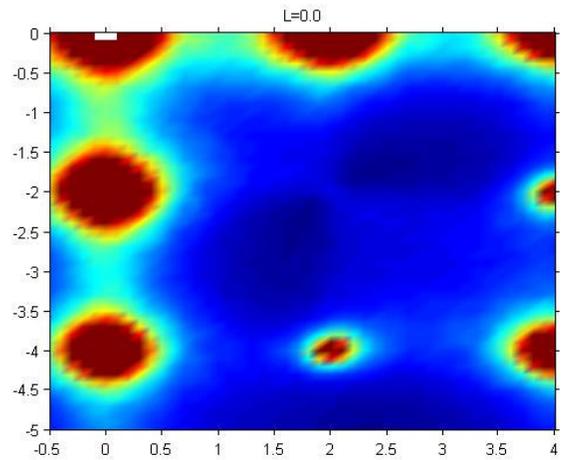

S3.

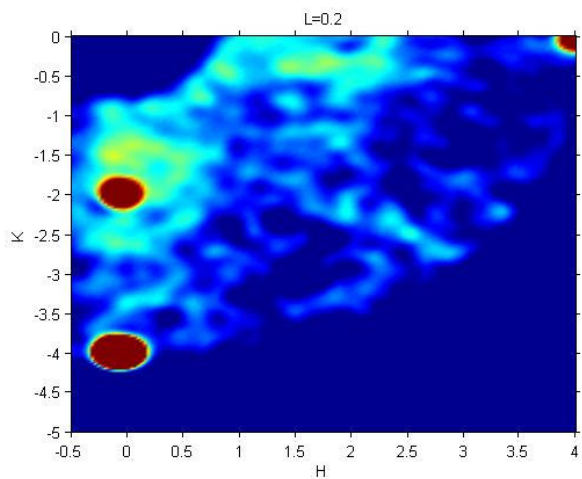

S4.

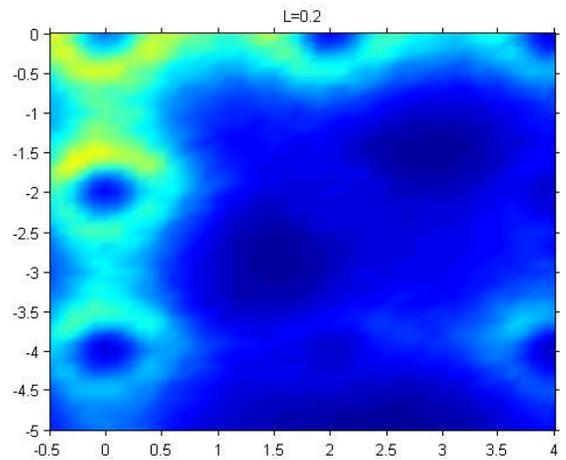

S5.

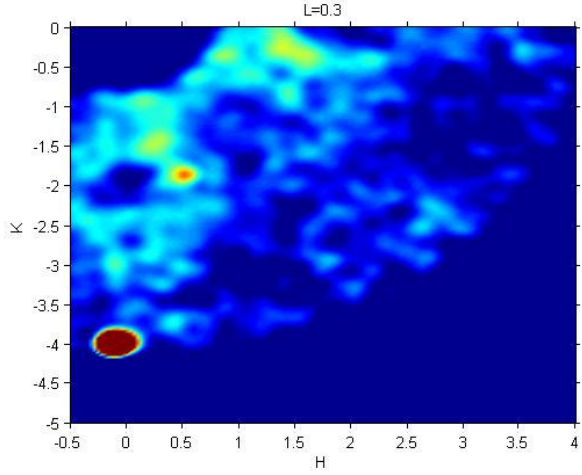 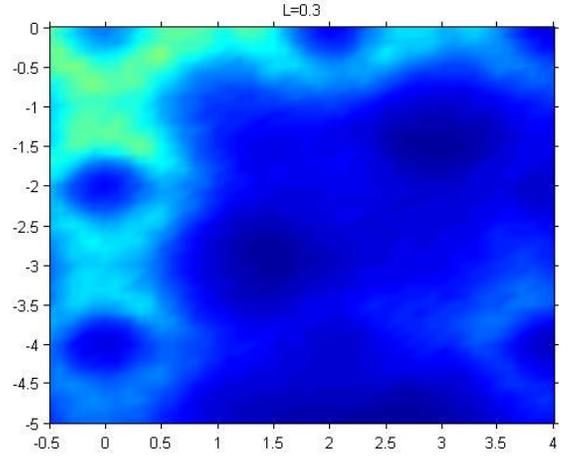

S6.                                        S7.

**Scattering calculations**

Spin-ice is equivalent to the antiferromagnetic Ising pyrochlore magnet when the spin configurations of one tetrahedral sublattice are mapped from *up* to *out* and *down* to *in*. Compared to the Ising solution, the neutron scattering cross-section is more complicated because it is non-collinear in the spin ice case and the off-diagonal components of the cross section are not identically zero. The cross section in terms of *in-out* variables $\mu_{i,l}$ on the four sublattices *i* of tetrahedra *l* are written in terms of the relevant correlation functions between sublattices *l,l'* $\langle \mu_l(\mathbf{k})\mu_{l'}(-\mathbf{k})\rangle$ as

$$\frac{d\sigma}{d\Omega} \propto |f(\mathbf{k})|^2 \sum_{l,l'}\sum_{\alpha,\beta}\left(\delta_{\alpha,\beta}-(k_\alpha k_\beta)/k^2\right)\varepsilon_l^\alpha \varepsilon_{l'}^\beta \langle \mu_l(\mathbf{k})\mu_{l'}(-\mathbf{k})\rangle \qquad (S1)$$

where α and β are Cartesian coordinates *x*, *y*, and *z* which also correspond to the cubic crystal directions **a**, **b**, and **c**. Here the definition of the unit vectors $\hat{\varepsilon}_l$ and the corresponding sublattices are given in the Table S1; note that the unit-length spin vector on the spin ice lattice is $\mathbf{S}_{i,l} = \mu_{i,l}\hat{\varepsilon}_l$.

| Sublattice | Position from centre of tetrahedra | Unit vector $\hat{\varepsilon}_l$ |
|---|---|---|
| 1 | $[-1,-1,-1]a/8$ | $[-1,-1,-1]/\sqrt{3}$ |
| 2 | $[1,1,-1]a/8$ | $[1,1,-1]/\sqrt{3}$ |
| 3 | $[1,-1,1]a/8$ | $[1,-1,1]/\sqrt{3}$ |
| 4 | $[-1,1,1]a/8$ | $[-1,1,1]/\sqrt{3}$ |

Table S1.

Now the correlation function $\langle \mu_l(\mathbf{k})\mu_{l'}(-\mathbf{k})\rangle$ is equivalent to that for the antiferromagnetic Ising pyrochlore *i.e.* $\langle \mu_l(\mathbf{k})\mu_{l'}(-\mathbf{k})\rangle = \langle S_l(\mathbf{k})S_{l'}(-\mathbf{k})\rangle$. These are calculated using the solution in Refs. *S4* and *S5*.

**Random walk correlation function**

The string geometry can be modelled as the path of a random walk where the spatial z-direction assumes the role normally played by time: with every step in the positive z-direction, the walker chooses between one of the two possible sites in the next xy-plane. Due to the cubic symmetry of the pyrochlore lattice, the x and y displacements appear symmetrically. This fixes the correlation function to be of Gaussian form characteristic of diffusive motion, yielding C(*x*,*y*,*z*), as in the text, where the constant $\gamma$ is related to the appropriate diffusion constant.

Random walks are further discussed in Ref. *S6*.

**Demagnetization effects**

Internal magnetic flux density, $\boldsymbol{B}_{int}$, was calculated using the finite element simulation software COMSOL Multiphysics to check for demagnetisation effects. The magnetisation value for the experimental temperature and applied magnetic field, along the [001] direction, was used with an input for a cylinder of geometry approximating the sample geometry used in neutron experiments. Initial measurements with the cylindrical axis at 45° to the field provided significant tilting of $\mathbf{B}_{int}$ away from [001] in agreement with the Boltzmann factors required to describe the diffraction data. The second experiment geometry was for the cylindrical axis to lie 90° to the applied field and allows $\boldsymbol{B}_{int}$ to be parallel to the required direction to within a few degrees. The cubic structure means that both geometries provide a [001] crystallographic direction to be parallel to the applied field. The calculated $\boldsymbol{B}_{int}$ value then went into Eqn. 1 ($\boldsymbol{h}=gm\boldsymbol{B}_{int}$) to find the biasing of the random walk.

**Magnetisation measurements**

Measurements of the magnetisation as a function of temperature and field were undertaken in St Andrews to complement the neutron measurements. The measurements were made using a home-built Faraday balance. The sample sits on a moving stage, suspended over four phosphor-bronze wires of calibrated elastic constant. We measure the magnetisation applying a known gradient of magnetic field and determining capacitively the distance between the moving stage and a fixed plate; this is converted into force (through the known elastic constants), and then into magnetisation (through the known gradient of field). As an initial check of the setup we measured a well know superconductor and a well known metamagnet. As a further test of our magnetisation setup we took high temperature magnetisation as a function of field (at 1.8 K) which we compared with identical measurements of the same sample taken with a Quantum Design SQUID. The magnetometer was attached to a dilution fridge, and thermal conduction to the sample was achieved through a set of silver wires directly attached to the mixing chamber of an Oxford Kelvinox 25 dilution refrigerator. The temperature of the sample was measured directly using a bare-chip $RuO_2$ thermometer attached to it.

**Heat Capacity measurements**

The specific heat was measured on a purpose-built calorimeter at the Helmholtz Zentrum Berlin (Laboratory for Magnetic Measurements) using a relaxation method. A single crystal of mass 10.31 mg (~0.8 x 2.8 x 0.6 mm³) was mounted on a sapphire calorimeter platform in a ³He-cryostat (*S7*). Data were taken for both warming and cooling in the temperature range 0.3 K to 4 K. This measurement was performed with zero applied magnetic field and the data from this are shown in Fig. 1B (open squares). Heat capacity and thermal conductivity data with applied field will be reported later by B.K. *et al*.

**Heat Capacity theory**

We have computed the heat capacity of a gas of magnetic monopoles with a Coulomb interaction in the framework of a Debye-Huckel theory (blue line in Fig. 1B). The interaction strength (given by the size of the monopole's magnetic charge, $Q = \frac{2\mu}{a_d} = 4.28 \times 10^{-13}$ J/Tm) and the bare monopole cost ($\Delta = \frac{2J}{3} + \frac{8}{3}\left(1+\sqrt{\frac{2}{3}}\right)D = 4.35$ K) were inferred from Ref. *S8* using parameters for the interactions in the range suggested by the detailed Monte Carlo-based microscopic modelling of Refs *S9* and *S10*. Debye-Huckel theory is a mean-field theory for mobile charges interacting via the Coulomb potential (*S11*). It allows us to obtain an approximate expression for the free energy of the system as a function of temperature, charge and chemical potential of the monopoles (*S12*), and from this we arrive at the specific heat using appropriate thermodynamic relations.

The theory breaks down when typical length scales (screening length, monopole separation) become of the order of the lattice constant. This happens for temperatures above approximately 1K, where the spin ice state gives way to a more conventional paramagnet, and monopoles cease to be suitable quasiparticles. Below this temperature, -- in particular for T<0.8K -- the theory works very well. Such good agreement cannot be achieved by a theory based on independent defects (as in a Bethe lattice calculation), even when allowing the particle creation cost to appear as a completely unrestricted fitting parameter. For comparison, we computed the specific heat of a single tetrahedron by explicit summation of the partition function over all $2^4$ states, using an effective nearest neighbour interaction $J_{eff}$. The result is illustrated by the red line in Fig. 1B, where we varied $J_{eff}$ to obtain the best agreement with the experimental data in the region $0.4$ K $< T < 1$ K.

The good low-temperature agreement of Debye-Huckel theory and experiment reinforces the suggestion that the energy in zero field equilibrates at temperatures lower than the magnetisation in field-sweep experiment: similarly, in zero field, muon experiments do detect dynamical ground states down to temperatures way below 600mK (*S13*,*S14*). The detailed origin, and the precise scope, of this phenomenon are as yet not properly understood.

**Non-interacting versus hard-core string model**

In the strong magnetic field limit, the magnetisation saturates and all spins are aligned by the field. In this limit, there are no monopoles, and hence no Dirac strings between them. We find that the diffuse scattering sharpens as the tilted field is increased at fixed temperature. In the tilted magnetic field case intensity is found in the diffuse scattering around positions that suggest some short-range hexagonal order (Figs. 4C and 4D), no such effects have been observed in the case where the field is parallel to the crystallographic axis. This may be due to the fact that the long-range interactions between the monopoles at the end of the strings become relatively more important as the paths of the strings become more spatially constrained, although there is as yet no detailed theory for this effect.

As one leaves the regime of low string densities, a `hard-core' contact interaction of the strings comes into play: when two strings meet, their combined entropy is lowered with respect to the entropy of two individual strings. In the "random walk with no interactions" model, independent random walks of spin flips are created and then S(q) for each walk is calculated and summed. Therefore the strings are not aware of other strings already existing, and the contact interaction is implicitly neglected. For the "hard-core interaction model", a string entering a tetrahedron which already has another string passing through it only has a unique `choice' where to exit. The structure factor S(q) is calculated once all of the strings have been created according to these rules. These two S(q) calculations provide similar agreement with data.

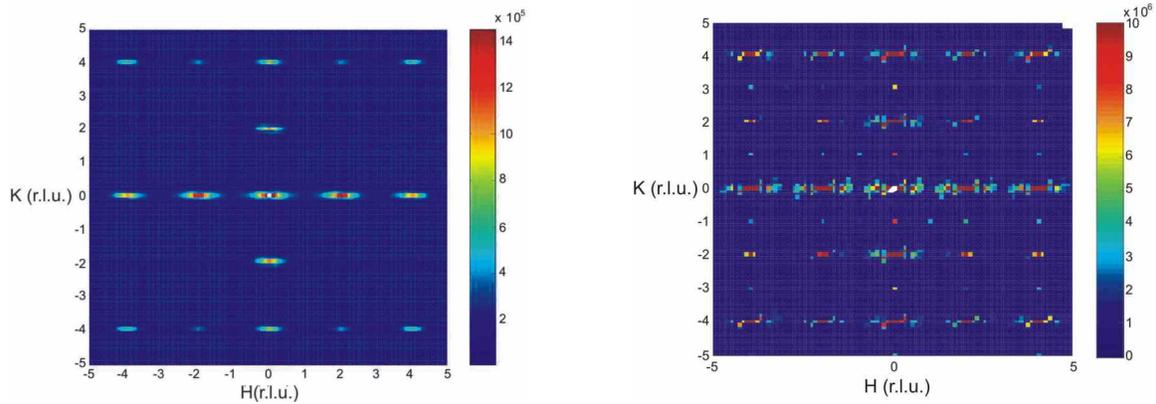

L=0 plane. Fig. S8) Left is the figure from the non-interacting string model. Fig. S9) Right is the hard-core string calculation with string density = 0.9. The diffuse scattering looks the same for both models.

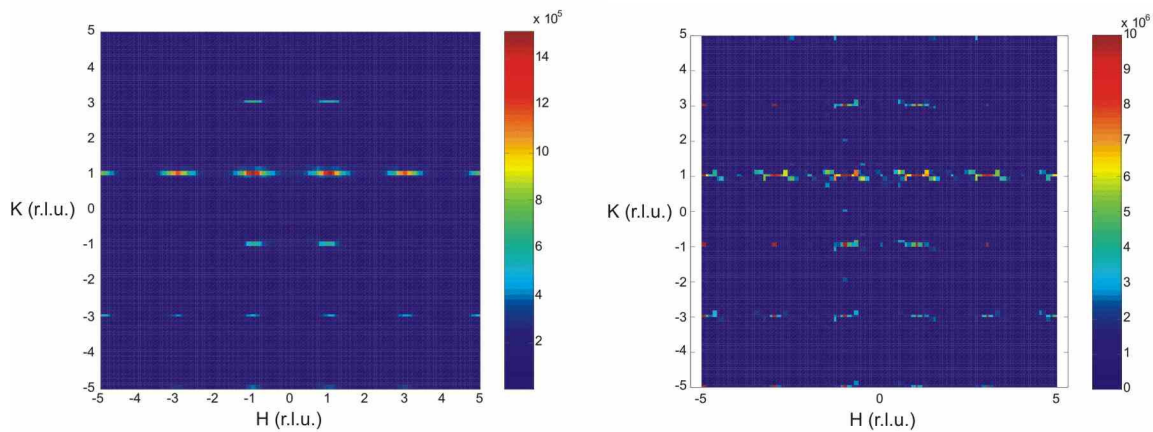

L=1 plane. Fig. S10) Non-interacting string model. Fig. S11) The figure on the right has a string density of 0.7 – scattering at integer *hkl* positions is a lot lower in intensity.

**Demagnetisation**

Demagnetisation effects which change the affective magnetic field within a sample are an important, yet sometimes overlooked, issue. Researchers would ideally choose a spherical sample shape because this has equal demagnetisation effects in the three dimensions of space. If the magnetisation of the material is low then the demagnetisation effect will be weak, and the sample dimensionality is not a big issue.

In $Dy_2Ti_2O_7$ we have a system with large magnetisation and therefore the demagnetisation is important. We utilised this effect to study tilted internal fields and their effect on the system.

Using COMSOL Multiphysics computer package enables the calculation of the internal magnetic flux density, $\vec{B}_{int} = \vec{B}_{app} + (1-N)\vec{M}$, which is the property to which the spins interact via the energy $-\vec{\mu} \cdot \vec{B}_{int}$, where $\vec{\mu}$ is the magnetic moment, $N$ is the demagnetisation factor, $\vec{M}$ is the magnetisation.

Below is shown the angle between $\boldsymbol{B}_{int}$ and the [001] direction and probability p(3→2) for a cylindrical sample with symmetry axis at 45° to the applied field. This is the geometry used in our tilted magnetic field measurements.

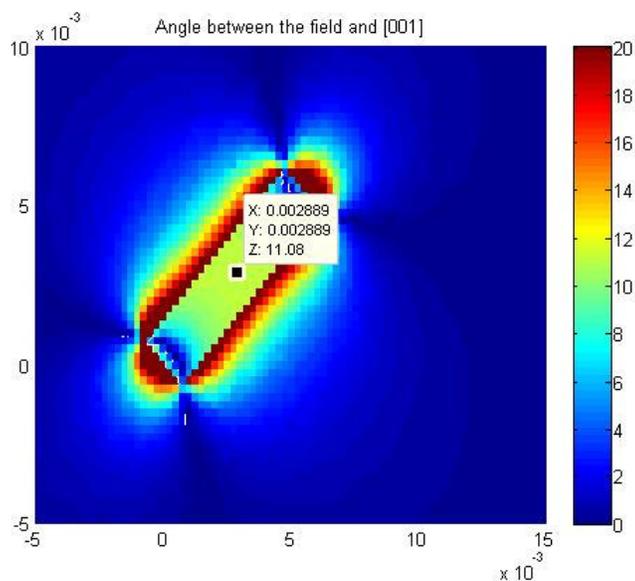

Fig. S12) Angle between $\bm{B}_{int}$ and [001] crystallographic direction. The x and y axis are distances in metres. The value in the box shows the angle (11.08°) at that position inside the sample.

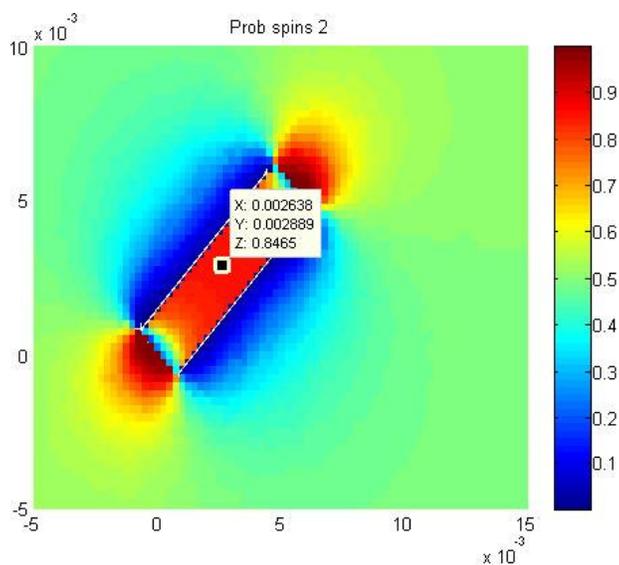

Fig. S13) Probability for random walk to move from site 3 or 4 to site 2, p(3→2). We found good agreement with p(3→2)=0.8 which is close to this maximum of 0.84.

**Supporting Information References**